\documentclass[prd,aps,preprintnumbers,twocolumn,floatfix,nofootinbib]{revtex4}
\usepackage{epsfig}
\usepackage{amsmath}

\def\eqn#1{eq.~(\ref{#1})}

\def\fig#1{fig.~{\ref{#1}}}

\def\M{{\mathcal M}}

\def\as{\ensuremath{\alpha_{s}}}
\def\a0{\alpha_0}

\def\ib{{\bar\imath}}
\def\vep{\varepsilon}
\def\lr{\leftrightarrow}

\def\Neqfour{{{\cal N}=4}}

\def\e{\epsilon}

\def\bea{\begin{eqnarray}}
\def\eea{\end{eqnarray}}

\def\be {\begin{equation}}
\def\ee {\end{equation}}

\begin{document}

\preprint{SLAC--PUB--13517} 

\renewcommand{\thefigure}{\arabic{figure}}

\title{Matter Dependence of the Three-Loop Soft Anomalous Dimension Matrix}

\author{Lance J. Dixon}

\affiliation{SLAC National Accelerator Laboratory, Stanford University,
  Stanford, CA 94309, USA}
\date{\today}

\begin{abstract}
The resummation of soft gluon exchange for QCD hard scattering
requires a matrix of anomalous dimensions, which has been
computed through two loops.  The two-loop matrix is proportional
to the one-loop matrix.  Recently there have been proposals 
that this proportionality extends to higher loops.
One can test such proposals by computing the dependence of this matrix 
on the matter content in a generic gauge theory.
It is shown that for the matter-dependent part the proportionality
extends to three loops for arbitrary massless processes.
\end{abstract}

\maketitle

High-energy colliders, particularly hadron colliders such as the Tevatron
and the imminent Large Hadron Collider (LHC), copiously produce
complex events containing multiple jets of hadrons.  The production rate
for such events can be understood through perturbative 
quantum chromodynamics (QCD).  However, in many cases large
logarithms of ratios of kinematic quantities appear,
and it is necessary to resum these logarithms to obtain
an accurate prediction.   For processes in which only 
two partons appear at the leading order in the perturbation expansion 
in $\alpha_s$ (Born level),
such as quark-antiquark annihilation to an electroweak vector boson,
or gluon-gluon fusion into the Higgs boson, such resummations
have been carried out to relatively high order~\cite{MVV05},
thanks in part to a detailed understanding of properties of the
Sudakov form factor in QCD~\cite{Sudrefs,Collins89,magnea90}.
In these cases, only one color structure is permitted --- the colors
of the two partons must be identical.  For three partons as well,
a unique color structure appears at Born level (an $SU(3)$
generator matrix or structure constant).

However, for four or more partons, such as appear in di-jet
production at hadron colliders, the Born level process contains
multiple color structures.  The emission or virtual exchange
of soft gluons can mix these structures.  To organize the logarithms
from virtual exchange it is convenient to introduce a 
{\it soft (or cross) anomalous dimension matrix}~\cite{%
BGSN,matrixdim,KorchemskyRadyushkin,KOS,KOSjet,BSZ}.
This matrix can be computed from the renormalization properties
of vacuum matrix elements of products of Wilson lines, 
also known as eikonal lines.  Each line corresponds to an external parton 
in the amplitude.  At one loop, 
the soft matrix can only change the colors of two hard partons 
at a time, because the virtual gluon propagator has only two ends.
Such action can be written as a color operator of the form
${\bf T}_i^a {\bf T}_j^a 
\equiv {\bf T}_i \cdot {\bf T}_j$~\cite{catani96,catani98}
where ${\bf T}_i$ indicates the action on the color of line $i$,
and the adjoint index $a$ of the virtual gluon is summed over.

At higher loops, it might seem that arbitrarily many external
partons (eikonal lines) 
could eventually be connected by soft gluons, resulting
in a very complicated soft anomalous dimension matrix.
For example, at two loops the structure 
$f^{abc} {\bf T}_i^a {\bf T}_j^b {\bf T}_k^c$ might have
appeared, connecting eikonal lines $i$, $j$ and $k$.  
Rather remarkably, however, such a structure does not appear,
for any two-loop amplitude in a massless gauge theory~\cite{ADS1,ADS2}.
Also, the two-loop soft anomalous dimension matrix ${\bf \Gamma}_S^{(2)}$
is proportional to the one-loop matrix,
\begin{equation}
{\bf \Gamma}_S^{(2)} 
= \Bigl[ \gamma_K^{(2)} / \gamma_K^{(1)} \Bigr] \, {\bf \Gamma}_S^{(1)}\,,
\label{Gamma2}
\end{equation}
where $\gamma_K^{(L)}$ is the $L$-loop coefficient
of the cusp anomalous dimension~\cite{KorchemskyRadyushkin},
$\gamma_K(\alpha_s) = \sum_{L=1}^\infty \gamma_K^{(L)} (\alpha_s/\pi)^L$,
with $\gamma_K^{(1)} = 2C_i$ ($C_i$ is the quadratic Casimir for line $i$).
This result agrees with explicit infrared singularities
of various two-loop QCD amplitudes~\cite{BDD03,glover04,BDK04}.
The soft anomalous dimension matrix 
also appears in the study~\cite{jantzen05} and 
resummation~\cite{manohar} of electroweak Sudakov logarithms,
relevant for very high-transverse-momentum processes at the LHC.
Recently it has been conjectured that the proportionality~(\ref{Gamma2})
might persist to higher loops~\cite{BCDJR,BecherNeubert,GardiMagnea}.
Ref.~\cite{GardiMagnea} also showed that such an ansatz is consistent
with a required invariance under rescaling of Wilson-line velocities,
but the solution to the consistency conditions is not unique
for amplitudes with four or more partons.

In this letter the proportionality~(\ref{Gamma2}) 
is shown to extend to three loops, for the part of ${\bf \Gamma}_S^{(3)}$ 
that depends on the matter content, in any massless gauge theory.
The result is established using the same transformation
of loop-integration variables employed at two loops in
refs.~\cite{ADS1,ADS2}.   There is also a term in 
${\bf \Gamma}_S^{(3)}$ arising from pure gluon exchange.
Although the result here shows that certain pure-glue 
contributions connecting three eikonal lines also vanish, 
other contributions connecting three and four lines remain to be
understood.  On the other hand, these contributions are the same 
in any theory, so it is possible, for example, to extract them 
from an amplitude in $\Neqfour$ super-Yang-Mills theory (SYM), 
and then apply the result to QCD.
In fact, the full color dependence of the three-loop four-point 
amplitude in $\Neqfour$ SYM is known in terms of a handful of 
loop integrals~\cite{BCDJR}.
Evaluating the integrals (a nontrivial task) 
would allow a test of the proportionality 
of the pure-gauge part of the soft anomalous dimension matrix 
at three loops.  The observed uniform {\it transcendentality}
of $\Neqfour$ SYM amplitudes~\cite{KLBCDKS} already implies that
any non-proportional term in ${\bf \Gamma}_S^{(3)}$ should be a 
weight-5 transcendentality function of kinematic invariants,
where $\zeta(n)$, $\ln^n x$ and ${\rm Li}_n(x)$ are examples
of weight $n$ functions --- in addition to the constraints of 
ref.~\cite{GardiMagnea}.

Consider the scattering amplitude for $n$ massless partons
$f_i$ carrying (all-outgoing) momenta $\{ p_i \}$ and color $\{r_i\}$.
To analyze the amplitude at fixed angles and
large overall momentum scale $Q$, 
represent the momenta as $p_i = Q v_i$, \quad $v_i^2=0$,
where the $v_i$ are four-velocities.  Divergences
are regulated using dimensional regularization with $D=4-2\e$.
The amplitude's color-dependence can be represented
as a vector $|\M\rangle$ in a $C$-dimensional vector space spanned by
the color tensors 
$\{ \left(c_I\right)_{\{r_i\}} \}$~\cite{KOS,catani98,TYS},
\begin{equation}
\label{amp}
\left|\, \M \right\rangle
\equiv 
\sum_{I=1}^C \M_{I} \, \left(c_I\right)_{\{r_i\}}
\,.
\end{equation}
For example, for $n$-gluon amplitudes, one choice
of basis is the set of multiple traces of generators,
${\rm tr}(T^{a_1}\ldots T^{a_k})
\,{\rm tr}(T^{a_{k+1}}\ldots T^{a_l})\,\ldots$.
For amplitudes with two quarks (1 and 3) and two anti-quarks (2 and 4), 
the basis is much simpler; it is spanned by 
$\delta_{i_1}^{\ib_2} \delta_{i_3}^{\ib_4}$
and $\delta_{i_1}^{\ib_4} \delta_{i_3}^{\ib_2}$.

Infrared divergences of on-shell amplitudes may be factorized
into jet functions describing virtual partons collinear
with the external lines, and soft functions characterizing the
exchange of soft gluons between the hard lines.
The general form of the factorized amplitude, for
equal factorization and renormalization scales $\mu$, 
is~\cite{TYS} 
\begin{eqnarray}
\label{facamp}
\left |\, {\cal M}
  \left(v_j,Q^2/\mu^2,\as(\mu),\e \right)\right \rangle
&=&\prod_{i=1}^{n} J^{[i]}\left(\as(\mu),\e
\right) 
\nonumber\\
&\ & \hspace{-45mm}
\times\, {\bf S}\left( v_j,Q^2/\mu^2,\as(\mu),
                     \e \right) 
\left|\,H\left(v_j,Q^2/\mu^2,\as(\mu)\right) \right \rangle
\,,~~
\end{eqnarray}
where $J^{[i]}$ is the jet function for external state $i$, 
${\bf S}$ is the soft function, and $H$ is the 
hard (short-distance) function, which is finite in the infrared.

The jet function for parton $i$ can be expressed to all orders in terms
of three anomalous dimensions, ${\cal K}^{[i]}$, ${\cal G}^{[i]}$
and $\gamma_K^{[i]}$; and ${\cal K}^{[i]}$ is completely determined
by $\gamma_K^{[i]}$~\cite{Collins89,magnea90}.
This letter concerns the soft function ${\bf S}$.
It is governed by the soft anomalous dimension matrix ${\bf\Gamma}_S$,
\begin{eqnarray}
&&{\bf S} \left(v_i\cdot v_j\,Q^2/\mu^2,\as(\mu),\e \right)
\nonumber\\
 &=&
{\rm P}~{\rm exp}\left[
\, -\; \int_{0}^{\mu} \frac{d\tilde{\mu}}{\tilde{\mu}}
{\bf \Gamma}_{S} \left(\frac{v_i\cdot v_j \,Q^2}{\mu^2} ,
  \bar\as\left(\tilde{\mu}, \e\right)\right) \right]
\nonumber\\ &=&
 1+\frac{1}{2\e}\left(\frac{\alpha_s}{\pi}\right){\bf \Gamma}_{S}^{(1)}\,
+\frac{1}{8\e^2}\left(\frac{\alpha_s}{\pi}\right)^2
\left({\bf \Gamma}_{S}^{(1)}\right)^2
\nonumber\\ && \hskip0.0cm\null
- \frac{\beta_0}{16\e^2}
\, \left(\frac{\alpha_s}{\pi}\right)^2{\bf \Gamma}_{S}^{(1)}\,
+\frac{1}{4\e}\,
\left(\frac{\alpha_s}{\pi}\right)^2
{\bf \Gamma}_{S}^{(2)} + \dots\,,~~~
\label{expoS}
\end{eqnarray}
and also involves the $D$-dimensional running coupling.
At one loop, $\bar\as$ is given by
\begin{equation}
\label{asinD}
\bar\as\left(\tilde{\mu},\e\right) =
\as(\mu) \left(\frac{\mu^2}{\tilde{\mu}^2} \right)^\e
\sum_{l=0}^{\infty}\left[\frac{\beta_0}{4\pi\e}
\left(\left(\frac{\mu^2}{\tilde{\mu}^2} \right)^\e-1\right) \as(\mu)
\right]^l
\nonumber
\end{equation}
with $\beta_0 = (11 C_A - 4 T_R n_f)/3$ in QCD.

The one-loop soft matrix is~\cite{KOS,catani98}
\bea
{\bf \Gamma}_{S}^{(1)}
&=& 
\frac{1}{2}\sum_{i=1}^n \sum_{j\ne i}\, {\bf T}_i\cdot {\bf T}_j
  \,\ln\left(\frac{\mu^2}{-s_{ij}}\right) \,,
\label{GammaS1T}
\eea
where $s_{ij} =(p_i+p_j)^2$.
Resummed cross sections are determined by the eigenvalues 
and eigenvectors of ${\bf \Gamma}_{S}$~\cite{KOSjet,BCMN,BSZ,DM05}.  
Proportionality of ${\bf \Gamma}_{S}^{(2)}$ to
${\bf \Gamma}_{S}^{(1)}$ implies that for fixed kinematics,
the same eigenvectors that diagonalize the one-loop matrix
also diagonalize the two-loop matrix, simplifying the integration over 
scale~\cite{ADS1,ADS2}.

\begin{figure}
\centerline{\epsfxsize=8.5cm \epsffile{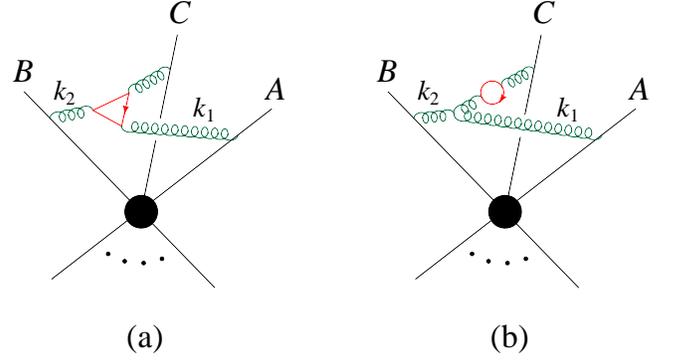}}
\caption{\label{from3gvfig}Three-loop diagrams containing a fermion
or scalar loop and three eikonal lines, corresponding to
a three-gluon vertex at two loops.}
\end{figure} 

\begin{figure}
\centerline{\epsfxsize=3.6cm \epsffile{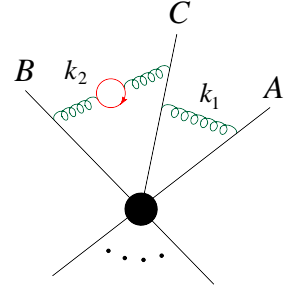}}
\caption{\label{from3Eno3gVfig}Three-loop three-eikonal
diagram containing a fermion or scalar bubble, and not
originating from a three-gluon vertex at two loops.}
\end{figure} 

The soft anomalous dimension matrix is found from the
single-ultraviolet-pole terms in $\e$, 
for suitable combinations of eikonal lines,
as described in detail at one~\cite{KOS} and two loops~\cite{ADS2}.
At $L$ loops, the maximum number of eikonal lines that can be
connected to some gluon is $2L$.  However, this maximal number 
corresponds to a disconnected product of $L$ one-loop single gluon
exchanges, which appears in the expansion of the 
exponential~(\ref{expoS}).
Thus the divergence is removed by one-loop renormalization.
The maximum number of eikonal lines that does not correspond to
a disconnected product of lower-loop configurations is $L+1$.
Requiring dependence on the matter content of the theory
means that there is a fermion or scalar loop in the diagram;
this cuts the maximum number of nontrivially connected legs
to $L$.  Such diagrams can be generated
from the maximal nontrivial $(L-1)$-loop configurations by inserting
bubbles with fermions or scalars, or triangles, or higher-point one-loop
diagrams.  

At two loops, the key eikonal diagrams for demonstrating proportionality 
involved three eikonal lines~\cite{ADS1,ADS2}, labelled by 
$A$, $B$ and $C$, with velocities $v_A$, $v_B$ and $v_C$.
The diagrams can be divided into two cases: 
({\it i}) three eikonal lines connected uniquely by a three-gluon vertex,
and ({\it ii}) three eikonal lines connected by two single-gluon
exchanges, so that two gluons attach to one of the eikonal lines.
Diagram ({\it i}) was found to vanish.  The sum of diagrams of type 
({\it ii}) was found to have only a symmetric color configuration,
proportional to 
$\{ {\bf T}_A \cdot {\bf T}_C\,,\, {\bf T}_C \cdot {\bf T}_B \}$;
the coefficient factorized into an appropriate form to
be removed by one-loop renormalization.  (It corresponds to
second-order terms in the expansion of the exponential~(\ref{expoS})
in which one leg is common to both factors.)

Inserting a matter triangle or bubble into diagram ({\it i})
gives the three-loop diagrams in \fig{from3gvfig}(a)
and (b).  Inserting a bubble into a type ({\it ii}) diagram
gives diagrams of the form shown in \fig{from3Eno3gVfig}.
I shall show that this insertion does not affect the
properties established for the pure-gluon diagrams at two loops.
The proof uses exactly the same change of integration variables 
used at two loops, plus, for diagram \ref{from3gvfig}(a), 
a handy representation of the one-loop three-gluon vertex 
$f^{a_1a_2a_3} \Gamma_{\mu_1\mu_2\mu_3}(k_1,k_2,k_3)$,
evaluated in an arbitrary gauge~\cite{DOT}.
Inserting this representation into diagram~\ref{from3gvfig}(a)
yields the loop-momentum integral,
\begin{eqnarray}
F_{1{\rm (a)}}(v_A,v_B,v_C)
&\propto&
f^{abc} {\bf T}_A^a {\bf T}_B^b {\bf T}_C^c 
\nonumber\\ &&\hskip-3.2cm \null
\times \int d^Dk_1 d^Dk_2\,  \frac{ \Gamma_{v_Av_Bv_C} }
{k_1^2 \, k_2^2 \, k_3^2 \, v_A\cdot k_1 \, v_B\cdot k_2 \, v_C\cdot k_3} \,,
 \label{F1a}
\end{eqnarray}
omitting the $+i\vep$ prescription, with $k_3 = -k_1-k_2$ and
\begin{widetext}
\begin{eqnarray}
\Gamma_{v_Av_Bv_C} &=& 
v_A^{\mu_1} v_B^{\mu_2} v_C^{\mu_3} \Gamma_{\mu_1\mu_2\mu_3}(k_1,k_2,k_3)
= A(k_1^2,k_2^2;k_3^2)\, v_A\cdot v_B \,v_C\cdot(k_1-k_2)
+ B(k_1^2,k_2^2;k_3^2) \,v_A\cdot v_B \,v_C\cdot(k_1+k_2)
\nonumber\\ &&\hskip0.0cm\null
- C(k_1^2,k_2^2;k_3^2) 
( k_1\cdot k_2 \, v_A\cdot v_B - v_A\cdot k_2 \, v_B \cdot k_1 )
\, v_C\cdot(k_1-k_2)
\nonumber\\ &&\hskip0.0cm\null
+ \textstyle{\frac{1}{3}} S(k_1^2,k_2^2,k_3^2) \, 
( v_A\cdot k_2 \, v_B\cdot k_3 \, v_C\cdot k_1
+ v_A\cdot k_3 \, v_B\cdot k_1 \, v_C\cdot k_2 )
\nonumber\\ &&\hskip0.0cm\null
+ F(k_1^2,k_2^2;k_3^2) \,
( k_1\cdot k_2 \, v_A\cdot v_B - v_A\cdot k_2 \, v_B \cdot k_1 ) \,
( k_2\cdot k_3 \, v_C\cdot k_1 - k_1\cdot k_3 \, v_C \cdot k_2 )
\nonumber\\ &&\hskip0.0cm\null
- H(k_1^2,k_2^2,k_3^2) [ v_A\cdot v_B
  ( k_2\cdot k_3 \, v_C\cdot k_1 - k_1\cdot k_3 \, v_C \cdot k_2 )
 - \textstyle{\frac{1}{3}}
    ( v_A\cdot k_2 \, v_B\cdot k_3 \, v_C\cdot k_1
    - v_A\cdot k_3 \, v_B\cdot k_1 \, v_C\cdot k_2 ) ]
\nonumber\\ &&\hskip0.0cm\null
 + \hbox{cyclic permutations of $(k_1,A)$, $(k_2,B)$, $(k_3,C)$.}
\label{vertex1l3g}
\end{eqnarray}
\end{widetext}
The precise forms of the coefficient functions $A$ through $H$ are given
in ref.~\cite{DOT}, for gluon and massless fermion loops.  Here only 
their symmetry properties are needed:  $A$, $C$ and $F$ are symmetric
under exchange of the first two arguments; $B$ is antisymmetric
in the first two arguments; $S$ is totally antisymmetric;
and $H$ is totally symmetric.  Ref.~\cite{DOT} does not explicitly 
consider the case of a massless scalar, but the vanishing
of three-point functions in $\Neqfour$ SYM implies that the scalar 
loop must have the same tensor decomposition and symmetry properties.

The change of loop-momentum variables used in refs.~\cite{ADS1,ADS2}
was expressed in terms of light-cone variables using the vectors
$v_A$ and $v_B$ to define $+$ and $-$ directions.  Here it suffices
to give the action of the transformation on the various Lorentz products
that appear,
\bea
v_A\cdot k_i &\to& \frac{v_A\cdot v_C}{v_B\cdot v_C}
                   \, v_B\cdot k_{\hat\imath} \,,
\qquad v_C\cdot k_i \to v_C\cdot k_{\hat\imath} \,,
\nonumber\\
v_B\cdot k_i &\to& \frac{v_B\cdot v_C}{v_A\cdot v_C}
                   \, v_A\cdot k_{\hat\imath} \,,
\qquad
k_i\cdot k_{j}  \to k_{\hat\imath} \cdot k_{\hat\jmath} \,,
\label{transfaction}
\eea
where $\hat{1}=2$, $\hat{2}=1$, and $\hat{3}=3$.
The Jacobian for this transformation is unity.
It is simple to check that every contracted tensor appearing
in \eqn{vertex1l3g} has the proper symmetry under~(\ref{transfaction})
so that, combined with the symmetry of the appropriate coefficient function,
and the remaining (symmetric) product of propagators in \eqn{F1a},
the integrand is antisymmetric.
Hence the integral~(\ref{F1a}) vanishes.\footnote{Another argument
for the vanishing uses the total antisymmetry of the integrated result
under exchanges of $\{ v_A, v_B, v_C \}$, plus invariance under 
suitable velocity rescalings~\cite{Private}.}
For example, the two terms multiplying
$S$ in \eqn{vertex1l3g} exchange with each other with a positive sign,
whereas the corresponding terms multiplying $H$ do so with a negative sign.
Note that the transformation~(\ref{transfaction})
is only to be used for the term shown explicitly in \eqn{vertex1l3g}.
For the two images of this term under cyclic permutation, 
one should use the correspondingly permuted transformation, 
which respects the symmetry properties of the permuted
functions $A$ through $H$.

For the diagrams in \fig{from3gvfig}(b) and \fig{from3Eno3gVfig}, 
write the bubble diagram as 
\begin{equation}
\Pi^{a_1a_2}_{\mu_1\mu_2}(k^2) = \delta^{a_1a_2}
 ( \eta_{\mu_1\mu_2} - \xi k_{\mu_1} k_{\mu_2}/ k^2 ) \Pi(k^2)
\,,
\label{propform}
\end{equation}
where $\xi=1$ for the transverse scalar and fermion bubbles,
but $\xi$ may differ from 1, depending on the gauge, for a 
gluon bubble.  The $\eta_{\mu_1\mu_2}$ term in \eqn{propform},
inserted into diagrams \ref{from3gvfig}(b) and \ref{from3Eno3gVfig},
clearly gives the same tensor structure as at two loops.
Thus~\cite{ADS1} its contribution to diagram \ref{from3gvfig}(b) vanishes.
It is easy to see that the $\xi$-dependent terms are also
odd under the transformation~(\ref{transfaction}),
so diagram \ref{from3gvfig}(b) vanishes for any $\xi$.

There are four diagrams of the type shown in \fig{from3Eno3gVfig},
related by changing the order of connection on line $C$, and
by moving the matter bubble to the other gluon line.
The sum of the four diagrams
splits into a color antisymmetric piece, proportional to
$[ {\bf T}_A \cdot {\bf T}_C\,,\, {\bf T}_C \cdot {\bf T}_B ]$,
whose coefficient vanishes exactly as in ref.~\cite{ADS2}; and
a color symmetric piece, proportional to 
$\{ {\bf T}_A \cdot {\bf T}_C\,,\, {\bf T}_C \cdot {\bf T}_B \}$,
containing a factor in the integrand of 
\begin{eqnarray}
 && \frac{1}{v_C\cdot k_1}
    \biggl[ v_A \cdot v_C 
   \biggl( v_B \cdot v_C - \xi \, \frac{v_B\cdot k_1 v_C\cdot k_1}{k_1^2}
    \biggr) \Pi(k_1^2)
\nonumber\\ &&\hskip0.0cm\null
   + (v_A \lr v_B, \ \ k_1 \lr k_2) \biggr]
 + (v_A \lr v_B, \ \ k_1 \lr k_2)
\nonumber\\
&=&   \frac{v_C\cdot (k_1+k_2)}{v_C\cdot k_1 \, v_C\cdot k_2 }
  \biggl[ v_A \cdot v_C 
   \biggl( v_B \cdot v_C - \xi \, \frac{v_B\cdot k_1 v_C\cdot k_1}{k_1^2}
    \biggr) 
\nonumber\\ &&\hskip0.0cm\null
\times \Pi(k_1^2)
  \  +\ (v_A \lr v_B, \ \ k_1 \lr k_2) \biggr] \,.
\label{ident2}
\end{eqnarray}
The latter form implies that this diagram factorizes
into the product of a one-loop diagram with the matter-dependent
part of the two-loop diagram.  Hence it also does not
contribute to ${\bf \Gamma}_{S}^{(3)}$.
The above arguments also show that, independent of the gauge,
the soft matrix does not receive contributions from those
three-eikonal pure-glue diagrams containing a gluon loop.

Finally one must address the matter-dependent contributions
to ${\bf \Gamma}_{S}^{(3)}$ involving only two eikonal lines.
As was the case for the full ${\bf \Gamma}_{S}^{(2)}$, 
only two of these diagrams (ladder and crossed ladder) 
do not obviously reduce group theoretically 
to the form ${\bf T}_i \cdot {\bf T}_j$.
Inserting a matter bubble of the form~(\ref{propform}) 
does not change the group theory.  The magnitudes of these 
two contributions are linked~\cite{cusp,cross}, however,
by the general non-abelian exponentiation
theorem for eikonal webs~\cite{nonabexp},
in such a way that their sum is of the form ${\bf T}_i \cdot {\bf T}_j$
(after removing the square of the one-loop term).
Because this result holds for the case of two partons ($i$ and $j$
of equal colors, so that ${\bf T}_i \cdot {\bf T}_j \to {\bf T}_i^2 =
C_i$), the constant of proportionality
must be the (matter dependent part of) the cusp anomalous dimension,
\begin{equation}
{\bf \Gamma}_S^{(3),\,{\rm matter}} 
= \Bigl[ \gamma_K^{(3),\,{\rm matter}} / \gamma_K^{(1)} \Bigr]
\, {\bf \Gamma}_S^{(1)}\,.
\label{Gamma3}
\end{equation}
An alternative, very general argument for the constant of 
proportionality, given that the matrix is proportional, is
based on an anomaly under rescalings of the 
eikonal-line velocities~\cite{GardiMagnea}.
As noted there, the situation may become more complex at four loops,
due to the fact that $\gamma_K^{(4)}$ might
contain quartic Casimir terms
$\sim(d_A^{abcd})^2$~\cite{nonabexp,cusp,AldayMaldacena}.
(At very strong coupling, ``quadratic Casimir scaling'' 
is known to break down for planar $\Neqfour$ SYM~\cite{Armoni}.)

In summary, the results of this letter show that the three-loop
soft anomalous dimension matrix is essentially the same in 
all massless gauge theories, {\it i.e.}
up to the matter dependence of $\gamma_K^{(3)}$.
If proportionality fails for the pure gauge terms, 
then at least the three-loop properties are constrained
by $\Neqfour$ super-Yang-Mills theory, as well as by the velocity
rescalings studied in ref.~\cite{GardiMagnea}.

\acknowledgments
This work was supported by the Department of Energy under 
contract DE--AC02--76SF00515.
I thank Adi Armoni, Thomas Becher, Zvi Bern, Gregory Korchemsky,
George Sterman, and especially Einan Gardi and Lorenzo Magnea,
for helpful correspondence.

\end{document}